# scientific reports

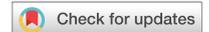

**OPEN**

# Single and multi-objective real-time optimisation of an industrial injection moulding process via a Bayesian adaptive design of experiment approach

Mandana Kariminejad[1,2,3], David Tormey[1,3], Caitríona Ryan[3,4,5], Christopher O'Hara[1,3], Albert Weinert[1,2,3] & Marion McAfee[1,2,3✉]

Minimising cycle time without inducing quality defects is a major challenge in injection moulding (IM). Design of Experiment methods (DoE) have been widely studied for optimisation of injection moulding, however existing methods have limitations, including the need for a large number of experiments within a pre-determined search space. Bayesian adaptive design of experiment (ADoE) is an iterative process where the results of the previous experiments are used to make an informed selection for the next design. In this study, an experimental ADoE approach based on Bayesian optimisation was developed for injection moulding using process and sensor data to optimise the quality and cycle time in real-time. A novel approach for the real-time characterisation of post-production shrinkage was introduced, utilising in-mould sensor data on temperature differential during part cooling. This characterisation approach was verified by post-production metrology results. A single and multi-objective optimisation of the cycle time and temperature differential ($\Delta T$) in an injection moulded component is proposed. The multi-objective optimisation techniques, composite desirability function and Nondominated Sorting Genetic Algorithm (NSGA-II) using the Response Surface Methodology (RSM) model, are compared with the real-time novel ADoE approach. ADoE achieved almost a 50% reduction in the number of experiments required for the single optimisation of $\Delta T$, and an almost 30% decrease for the optimisation of $\Delta T$ and cycle time together compared to composite desirability function and NSGA-II. The optimal settings identified by ADoE for multiobjective optimisation were similar to the selected Pareto optimal solution found by NSGA-II.

Injection moulding (IM) is one of the leading manufacturing processes for the mass production of complex-shaped plastic components that require high precision. The quality and efficiency of the process depend strongly on finding the optimum process settings, monitoring the process inline, and adapting the settings as needed to prevent defects in the component[1]. Non-uniform temperature distribution during the cooling stage of the injection moulded part is one of the main causes of part defects, leading to residual stresses, shrinkage, and warpage. For precision components, ensuring dimensional stability is crucial, requiring the elimination of shrinkage and warpage to meet the dimensional tolerances and prevent potential failures. Locally heated regions in the component, known as hotspot locations, must be identified and minimised to prevent residual stresses, which lead to the gradual manifestation of shrinkage and warpage in the days/weeks following production.

[1]Centre for Precision Engineering, Materials and Manufacturing Research (PEM Centre), Atlantic Technological University Sligo, Ash Lane, Sligo F91 YW50, Ireland. [2]Centre for Mathematical Modelling and Intelligent Systems for Health and Environment (MISHE), Atlantic Technological University, Ash Lane, Sligo F91 YW50, Ireland. [3]I-Form, Research Centre for Advanced Manufacturing, John Hume Institute, University College Dublin, Belfield, Dublin 4, Ireland. [4]TCD Biostatistics Unit, Discipline of Public Health and Primary Care, School of Medicine, Trinity College Dublin, Dublin, Ireland. [5]Wellcome-HRB Clinical Research Facility, St James's Hospital, Dublin D08 NHY1, Ireland. ✉email: marion.mcafee@atu.ie

Besides component quality, another critical parameter that directly affects the efficiency and cost of the process is cycle time. The cycle time of the IM process is the total time required to complete all the stages including mould closing, filling, packing, cooling, mould opening and ejection. Achieving the shortest possible cycle time while producing a product without defects is challenging. Several efforts have been made to minimise hotspots by redesigning cooling channels and runner systems and modifying the machine tools[2,3]. However, the complete elimination of hotspots through the tool design is rarely possible for complex parts, and it is important to find the process settings required for optimum temperature distribution, part quality and cycle time.

Various research have previously examined the problem of finding the optimum process settings to minimise defects or cycle time in industrial processes using different design of experiment (DoE) approaches. Classic design of experiment (DoE) approaches are commonly used to model the relationship between the process variables and desired responses. One of the popular DoE approaches used in industry is response surface methodology (RSM). One of the main categories of RSM is central composite design (CCD). For multi-objective optimisation, a classic DoE method needs to be combined with other optimisation approaches such as desirability function, Genetic Algorithm (GA), Artificial Neural Network, or Particle Swarm Optimisation (PSO), etc.

Mukras et al.[4,5] applied CCD along with other optimisation approaches to find the relationship between part quality factors and seven process parameters. In[4], the surface plot data from CCD was used to formulate an optimisation problem for finding the minimum cycle time with product shrinkage and warpage constraints. They solved the optimisation problem and validated their method via experiment. In[5], CCD was applied along with a GA to optimise a multi-objective problem that could achieve the minimum component warpage and shrinkage. They validated their approach via experiment, where a 7% error was found between the predicted optimal results and the actual experimental results. Zhijun et al.[6] established a multiobjective optimisation approach using CCD and GA to optimise warpage and sink marks in the seat belt cover plate of a car. The comparison of the theoretical and experimental values for warpage and sink marks indicated 0.2% and 7.95% error respectively.

Sudsawat and Sriseubsai[7] applied CCD combined with the firefly algorithm[8] to find the optimum process settings for minimising the warpage of an injection moulded component. Their method found suitable process parameters that reduced the component warpage by almost 23% compared to the recommended settings calculated by Moldex3D, (a commercial simulation software package for injection moulding), and empirical tests. Toh and Hassan[9] applied CCD to find the optimum mould temperature and packing pressure required to minimise the warpage and flash in a multi-cavity injection moulding process. The overlaid contour plots for the eight cavities and both responses led to finding the feasible region to minimise the response. The results were validated through physical experiments. A detailed review of recent studies applying classic DoEs with various optimisation techniques in IM since 2018 is listed in Table 1.

In these classic design of experiment approaches, the sampling and experiment patterns are deterministic and cannot change or adapt to the features that appear during the experiment. Typically, a large number of experiments are required for effective optimisation, and this may have to be repeated multiple times, for example on changing material batch or supplier. Also, employing robust and strong global optimisation techniques like firefly algorithm, MOPSO, MOGA, and NSGA-II alongside classic DoEs across diverse applications[10–13], often enhances model performance and optimisation effectiveness. However, unlike ADoE, these methods often lack efficiency in sampling the experimental space, do not support inline optimisation, and primarily rely on traditional DoE for data generation.

Adaptive design of experiment (ADoE), on the other hand, is an iterative process where the results of the previous experiment are used to make an informed selection for the next design. The adaptive approach has several advantages over classic DoEs like RSM. Firstly, it is tailored to the design process (incorporating prior knowledge). Secondly, the search and optimisation processes are rapid, and the number of experiments is reduced due to the exploitation of prior knowledge to reduce the complexity and accelerate the process. Thirdly, adaptive methods can search in a high-dimensional space, while classic DoEs have a limited search space. Finally, ADoE can be used in multi-objective optimisation problems[14]. ADoE has been enabled through a combination of machine-learning approaches applied to the experimental design. Bayesian optimisation (BO) is a useful tool for ADoE using adaptive sampling to find a global optimum for a response. It can be applied even when the objective function is unknown or complex, the evaluation of the objective function is expensive, and when the gradient information is unavailable[15,16].

| DoE method | Objective(s) | Optimisation approach | References |
|---|---|---|---|
| Orthogonal testing | Warpage, shrinkage | NSGA-II | [17] |
| Latin hypercube design | Warpage, shrinkage | NSGA-II | [18] |
| Box-Behnken | Max shrinkage, volume shrinkage, addendum & circle diameters shrinkage | NSGA-II | [19] |
| RSM | Warpage, residual stresses | PSO (simulation) | [20] |
| CCD | Warpage | Firefly algorithm | [7] |
| CCD | Cycle time | Nonlinear constrained optimisation | [4] |
| CCD | warpage, shrinkage | GA | [5] |
| CCD | Warpage, flash | Overlaid contour plots | [9] |
| CCD | Warpage, sink marks | GA | [6] |
| Taguchi | Warpage, shrinkage, residual stresses | Nondominated sorting genetic algorithm II (NSGA-II) | [21] |
| Taguchi | Surface roughness, shrinkage | Composite desirability function | [22] |
| Taguchi | Warpage, clamp force, weld line | ANN- MOGA | [23] |
| Taguchi | Cycle time, warpage | Composite desirability function | [24] |
| Taguchi | Shrinkage, warpage, short shot | Fuzzy decision-making process | [25] |
| Taguchi | Cycle time, weight, energy | ANN-MOGA | [26] |
| Taguchi | Mechanical properties, shrinkage | Multi-criteria decision making (MCDM) & NSGA-II | [27] |

**Table 1.** Previous studies on the optimisation of IM using classic DOEs since 2018.

ADoE and BO approaches have emerged from the initial application in medical science for dose-finding studies[28], to recent investigations in industrial applications, such as metal processing to optimise alloy composition and heat treatment[29,30], waste treatment process[31], and additive manufacturing to optimise metamaterial composite, lattice structure design and process parameters[32–35]. However, to date, there are few practical studies of ADoE in manufacturing and none in relation to polymer processing, despite recent efforts in multiobjective optimisation of such processes using traditional DoE methods with modern optimisation algorithms as outlined above.

There have been a few attempts using adaptive DoE for single response optimisation of the injection moulding process through simulation models. These studies[36–38] devised kriging and GP surrogate models to find the relationship between process parameters and warpage through simulation. The final optimum process parameters found with these approaches showed a good convergence with the experimental results of warpage. Recently, Jung et al. applied Bayesian optimisation for multi-objective optimisation of warpage and cycle time[39], also using Moldflow simulation. They compared the Bayesian optimisation method to a constrained generative inverse design network (GIDN), which uses backpropagation to find the gradient of an objective function. Simulation software was used for applying both methods. The results showed the efficiency of both approaches in multi-objective optimisation, while the Bayesian framework had a better performance in adapting to the variations in responses with a smaller dataset required.

The above studies employed data from simulation software to develop the surrogate model and optimise the part quality. However, basing the process optimisation on simulation software such as Moldflow and Moldex has several drawbacks. Each simulation is extremely time intensive (several hours), purchasing the software is costly, and the calculations are not precise since the parameters of the material are often not well defined or may even change from batch to batch. Hence an efficient experimental method of process optimisation is required to overcome these shortcomings.

The challenge in applying adaptive approaches experimentally is that the product quality in terms of dimensional stability is not known until metrology measurements have been obtained, usually a week or more post-production. Hence, in order to explore the potential benefits of an ADoE approach, a rapid indicator of dimensional stability is required. The part temperature distribution is a good indicator of the product quality since a uniform temperature profile eliminates thermally-induced residual stresses and hence shrinkage or warpage. This study proposes a real-time indicator of temperature profile for part quality evaluation in terms of the temperature differential between a known hotspot location in the component and a location with average temperature using two in-mould thermocouples.

This paper, for the first time, investigates the practical effectiveness of the Bayesian Optimisation ADoE compared to three well-established multi-objective optimisation methods within an injection molding process using in-line measurements. The objective is to find the optimum process settings that minimise both the cycle time and the temperature differential ($\Delta T$) between a part hotspot and an average temperature point, which has been validated in the study to correlate to post-production shrinkage. Process parameters were initially screened using fractional factorial DoE to identify significant factors. A Central Composite Design (CCD) was then employed to collect data and build quadratic models for each output. For single-objective optimisation of the temperature differential ($\Delta T$), both the desirability function and adaptive DoE (ADoE) were applied, with results compared for each method. For multi-objective optimisation of $\Delta T$ and cycle time, desirability function and Nondominated Sorting Genetic Algorithm (NSGA-II) were used, and ADoE was also compared against these methods for evaluating its overall effectiveness.

The main novelties of this study are (1) the introduction of a real-time monitoring method demonstrated to be a suitable proxy for post-production shrinkage measurements, and (2) validation of an intelligent adaptive design of experiment (ADoE) method for multi-objective optimisation in an injection moulding process using experimental data.

This research shows the suitability of the in-line method for predicting the dimensional stability of the part and highlights the superior performance of the ADoE approach. Unlike other multi-objective optimisation methods, which required 33 experiments to identify optimal objectives, the ADoE approach significantly reduced this number to just 22 for multi-objective optimisation of the temperature differential ($\Delta T$) and cycle time, and to 16 for single-objective optimisation of $\Delta T$. Additionally, the optimal process settings identified through ADoE led to enhanced part quality and reduced shrinkage, highlighting its potential for improving manufacturing outcomes.

The remainder of the paper is organised as follows. The theory section outlines the methods used in the study, including Bayesian Optimisation, CCD, optimisation by desirability function, and NSGA-II. The next section details the methodology, including experimental set-up, design, and data collection, and details on the optimisation methods applied. The corresponding results and discussion of the experimental studies are defined in the Results and Discussion section, followed by a conclusion with suggested further research.

## Theory
### Bayesian ADoE

Bayesian optimisation (BO) consists of two main components: a statistical and probabilistic surrogate model to estimate the black-box objective function, and an acquisition function to decide on the next sampling point[40]. A Gaussian process (GP) regression is typically considered the best choice for creating the surrogate model. It can estimate the values of the black-box function $f(x)$ at the point $x$, by providing a posterior probability distribution. $f(x)$ is Gaussian if any vector of $x = [x_1, x_2, ..., x_n]^T$ such that $x_i \in X$ (input space), the vector of output $f(x) = [f(x_1), f(x_1), ..., f(x_1)]^T$ is Gaussian distributed with mean $[m(x_1), m(x_1), ..., m(x_1)]^T$. The GP considers the prior distribution to be normal and is specified by a mean vector $\mu(x)$ and a covariance matrix $k(x, x')$ between $f(x)$ and $f(x')$ where $x$ and $x'$ are a pair of observations:

$$f(x) \sim Gaussian(\mu(x), k(x, x')) \tag{1}$$

The covariance matrix, also called a kernel, captures the smoothness of the process. If $x$ and $x'$ are close to each other, then their associated values for the output ($f(x)$ and $f(x')$) will also be close, and the closeness can be estimated by the distance between these two points. One commonly used kernel is the Matern kernel, defined in Eq. (2). $k(v)$ is a modified Bessel function, $\Gamma$ is the gamma function, $v$ (known as a smoothness parameter) and $\alpha_0$ are the hyperparameters.

$$k(x, x') = \alpha_0 \frac{2^{1-v}}{\Gamma(v)} (\sqrt{2v}||x - x'||)^v k(v)(\sqrt{2v}||x - x'||) \tag{2}$$

If the $v$ is a half-integer (1/2, 3/2, and 5/2), the kernel will be a product of an exponential and a polynomial function, as written in equation (3).

$$k(x, x')_{v=1/2} = \alpha_0 exp(-||x - x'||)$$
$$k(x, x')_{v=3/2} = \alpha_0 exp(-\sqrt{3}||x - x'||)(1 + 3||x - x'||)$$
$$k(x, x')_{v=5/2} = \alpha_0 exp(-\sqrt{5}||x - x'||)(1 + 5||x - x'|| + \frac{\sqrt{5}}{3}||x - x'||^2) \tag{3}$$

In a physical experiment, observations can generally be considered subject to normally distributed noise $\varepsilon \sim N(0, \sigma^2)$ and the objective function at point $x$ can be written as:

$$y = f(x) + \varepsilon \tag{4}$$

For the prediction of the objective function at the new point x, we consider $D_{1:n} = (x_i, y_i)$ and the result is a normal distribution with mean $\mu(x)$ and variance $\sigma_n^2(x)$, the objective function can be calculated based on equation (5).

$$p(f(x)|D_{1:n} x) = N(\mu_n(x), \sigma_n^2(x))$$
$$m_n(x) = m(x) + k(x, x_{1:n})\left[k(x_{1:n}, x_{1:n}) + \sigma_{noise}^2 I\right]^{-1}(y_{1:n} - \mu(x_{1:n}))$$
$$\sigma_n^2(x) = k(x, x) - k(x, x_{1:n})\left[k(x_{1:n}, x_{1:n}) + \sigma_{noise}^2 I\right]^{-1}k(x_{1:n}, x)$$
$$k(x, x_{1:n}) = [k(x, x_1), k(x, x_2), ..., k(x, x_n)]$$
$$k(x_{1:n}, x_{1:n}) = \begin{bmatrix} k(x_1, x_1) & \cdots & k(x_1, x_n) \\ \vdots & \ddots & \vdots \\ k(x_n, x_1) & \cdots & k(x_n, x_n) \end{bmatrix} \tag{5}$$

The second element of BO is an acquisition function (AF) which can be calculated from the mean and variance of the GP model, meaning it is rapid and inexpensive to compute. The acquisition function decides which area of the input domain maximises exploitation (where the objective mean is high) and exploration (where the uncertainty is high) for the next evaluation of the black-box function f(x). For example, the AF value is small in the areas where f(x) is suboptimal or has already been searched. The desired next experimental setting is where the AF is maximum. There are various types of AFs, and one commonly used choice is Expected Improvement (EI), proposed originally by Mockus[41]. The EI can be calculated as Eq. (6), where $f(x^*)$ is the current maximum, $\Phi$ is the cumulative distribution function (CFD), and $\phi$ is the probability distribution function (PDF) of the normal distribution to balance between exploration and exploitation. The following conditions should be met: first, $m_n(x) > f(x^*)$; second, the variance $\sigma_n(x)$ should be increased to maximise the EI based on this equation.

$$EI(x) = (m_n(x) - f(x^*))\Phi(\frac{\mu_n(x) - f(x^*)}{\sigma_n(x)}) + \sigma_n(x)\phi(\frac{\mu_n(x) - f(x^*)}{\sigma_n(x)}) \tag{6}$$

Figure 1 illustrates the BO process. A Gaussian Process surrogate model is fitted to some observations with the uncertainty (confidence intervals) shown by the green area. The red plot represents the fitted acquisition function based on the Gaussian Process model and illustrates that the AF is high in previously unsampled regions where the mean of the AF is also predicted to be high. The next sample point is chosen as the point where the AF is maximum (indicated red star in the graph). For further in-depth reading on the theory of Bayesian optimisation and adaptive design of experiments, the reader is referred to the following references[14,42–44].

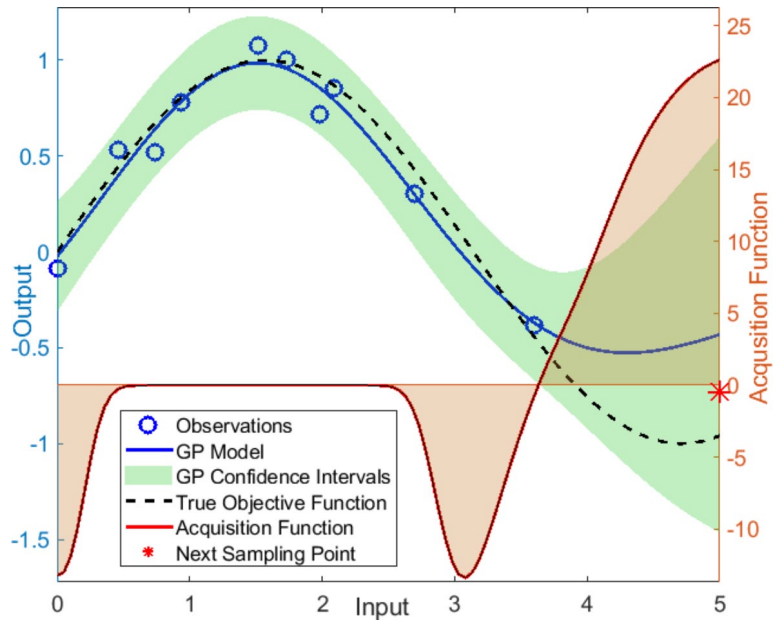

**Fig. 1.** Bayesian optimisation using a Gaussian process model and acquisition function for response minimisation.

## Optimisation using the desirability function

The desirability function, or multiple response optimisation, was developed by Derriger and Suich[45] and is a technique that transforms a multi-objective optimisation problem into a scale-free value known as desirability. This method integrates these individual objectives into a single optimisation problem, rather than solving them separately. The optimisation consists of two main steps. First, the value of each response is assigned to a dimensionless desirability $d_i$ that $0 < d_i < 1$. For values of $d_i$ near 0, the response is undesirable; on the contrary, the response for values of $d_i$ close to 1 is desirable. Secondly, each of the desirability values are combined into an overall desirability function, called D, that is calculated using Eq. (7), where n is the number of responses and $y_i$ is the desired response. A weight ($w_i$) can also be assigned to each response based on its effect on the final objective to adjust the desirability function as in Eq. (7).

$$D = \left(\prod_{i=1}^{n} d_i(y_i)\right)^{1/n} = [d_1(y_1) \times d_2(y_2) \times \cdots \times d_n(y_n)]^{1/n}$$

$$D = [d_1(y_1)^{w_1} \times d_2(y_2)^{w_2} \times \cdots \times d_n(y_n)^{w_n}]^{1/n}$$

(7)

In an optimisation problem, the desired response can be minimised, maximised, or set as a target value. A single response optimisation indicates how the input variables affect the desirability of the related response. In this research, the desired responses are cycle time and the temperature differential between a hotspot and a point of average temperature. The goal is to minimise these two values. The desirability function for minimisation with a weight equal to 1 (since both responses were considered to have the same importance on the quality of the component) for each of the responses is described by Eq. (8), where $y_{min}$ and $y_{max}$ are the minimum and maximum acceptable values for the response $y_i$[46]. The multi-objective optimisation by desirability function in this research was carried out using a response optimiser in statistical software Minitab 20.

$$d_i(y_i) = \begin{cases} 1 & if \quad y_i \leq y_{min} \\ \frac{y_{max} - y_i}{y_{max} - y_{min}} & if \quad y_{min} \leq y_i \leq y_{max} \\ 0 & if \quad y_i \geq y_{max} \end{cases}$$

(8)

## Non-dominated sorting genetic algorithm (NSGA-II)

NSGA-II is an evolutionary algorithm developed to overcome the shortcoming of classical multiobjective optimisation approaches which convert the optimisation problem into a single objective optimisation. It incorporates mechanisms to include elitism and preserve diversity. It has two further stages compared to GA: non-dominated sorting and crowding distance ranking. In non-dominated ranking individuals (chromosomes) are sorted into Pareto fronts based on their dominance. In the dominance relation, solution $p$ dominates solution $q$ (donated as $p \leq q$) if it is at least as good as solution $q$ and strictly better in at least one objective. In other words, a solution is Pareto optimal if it is not dominated by any other solution in the objective space[47]. A set of Pareto optimal solutions, also called the Pareto front, represents the trade-off between the conflicting objectives. Crowding distance estimates the density of the solutions around a specific solution in the objective space. It maintains diversity in the population and guides the search towards the Pareto front[48,49]. Figure 2 represents the principle of the NSGA-II algorithm.

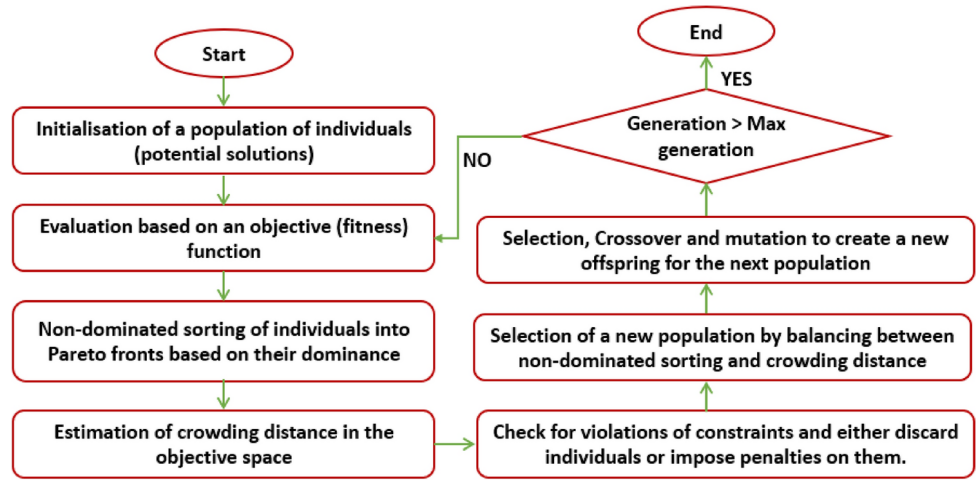

**Fig. 2**. Flowchart of NSGA-II optimisation algorithm.

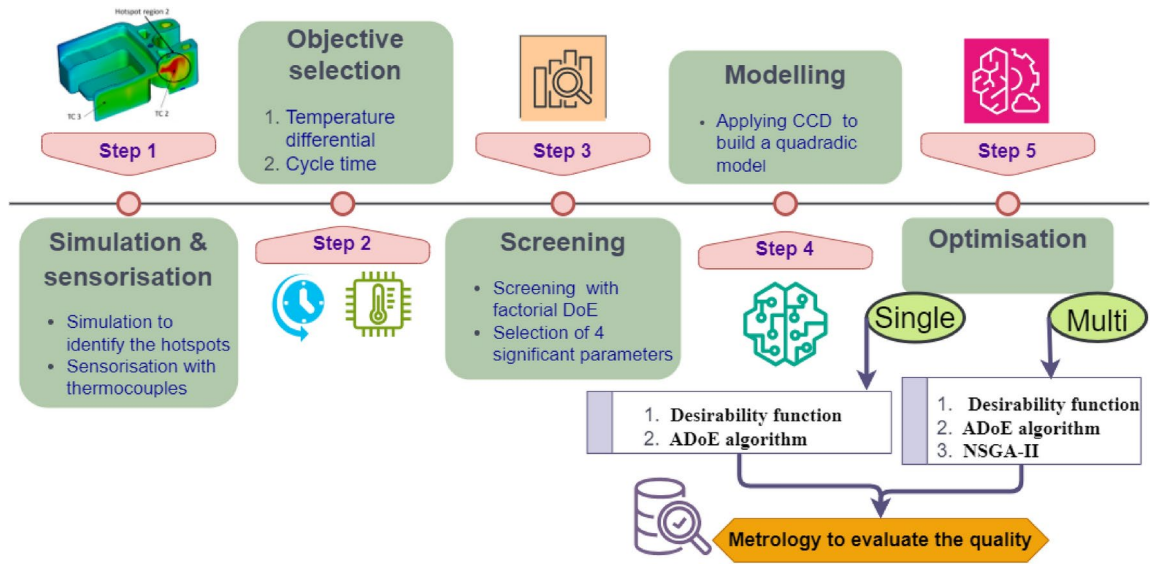

**Fig. 3**. Flowchart of the optimisation workflow.

## Methodology
This section outlines the experimental methodology, including: the simulation of the process; sensorisation; screening of the input parameters; implementation of CCD to build a quadratic model; single objective optimisation with the desirability function and ADoE; multi-objective optimisation with the desirability function, ADoE, and NSGA-II; and finally, the evaluation of quality by metrology. The overall procedure is illustrated in Fig. 3.

### Equipment and sensorisation
The isometric view and dimensions of the injection moulded part are shown in Fig. 4a. The material of the component is POM (Polyoxymethylene), and the Clip is currently being manufactured by an industrial partner with issues such as excessive cycle time and part shrinkage. In our previous research[50], the temperature distribution and hotspot regions were identified within the part through Moldflow simulation software as shown in Fig. 4b.

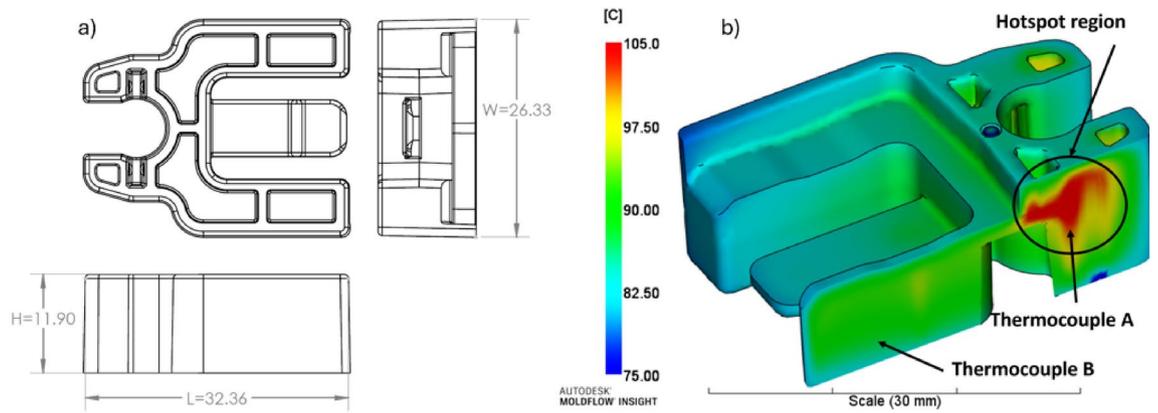

**Fig. 4.** (**a**) Clip main dimensions in millimetres. (**b**) Temperature distribution in part by simulation.

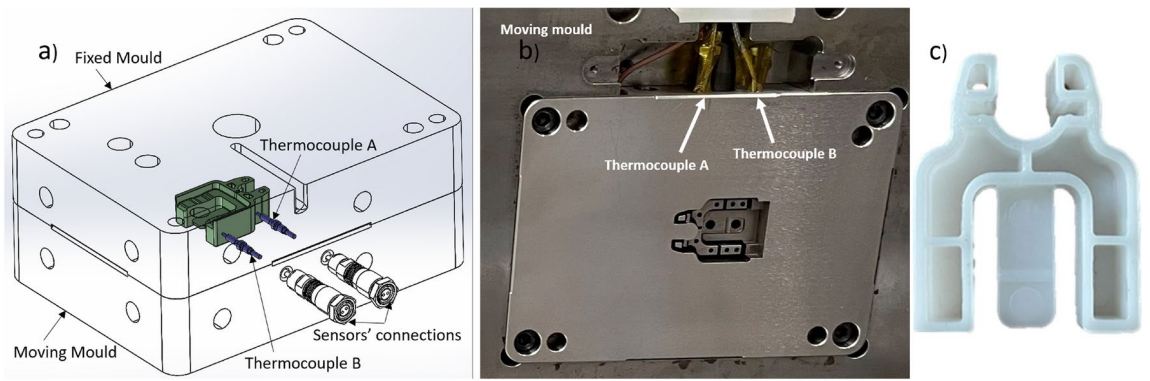

**Fig. 5.** (**a**) Isometric view of the sensors in the mould tools. (**b**) Manufactured mould tools and (**c**) the moulded clip.

Sensors are required to measure the temperature distribution in the part and extract accurate real-time data that can be used for optimisation. Kistler high-temperature thermocouples (type 6193B0,4) were selected for direct in-mould temperature measurement. The key selection criteria include the sensor temperature durability of up to $450^{\circ}C$, and the small size (diameter of 1 mm) needed to fit into the cavity.

Informed by the Moldflow simulation results, sensor locations were identified and two direct Kistler thermocouples were placed in the mould insert, one to take measurements of the identified hotspot point (Thermocouple A) and another to measure a point with average temperature (Thermocouple B) based on the simulation results in Fig. 4b. Figure 5 shows the 3D view, the manufactured moving half of the mould inserts with the embedded thermocouples and the manufactured product (Clip).

## Screening of the parameters and CCD implementation

This study aimed to optimise two responses. The first response of interest is the temperature differential ($\Delta T$). In the first part of the Results and Discussion section, it is validated that $\Delta T$ correlates with shrinkage and has been chosen as the quality index to update the model in real-time. $\Delta T$ can be measured by calculation of the difference between thermocouple A readings from thermocouple A (the hotspot location) and thermocouple B (the average temperature) in Figs. 4 and 5 and it can be written as follows:

$$\Delta T(^{\circ}C) = T_{TCA} - T_{TCB} \tag{9}$$

The second response is cycle time, which can be recorded from the control panel of the injection moulding machine for each cycle. The only process variables that influence the cycle time are cooling and packing (holding) time which can be defined in the machine settings. In comparison, there are many process parameters that will affect the temperature differential. If all of the input variables are considered, the number of experiments will be very high. Hence screening is essential to evaluate the main variables and their interaction effects on the desired outputs. Parameter selection for screening was driven by the primary settings of the process controller on the injection moulding machine, ensuring that the chosen parameters are directly relevant to the process outcomes. Initially, eight process parameters were identified for screening to assess the main design variables. Subsequently, insignificant parameters were removed based on Analysis of Variance (ANOVA), which was employed to identify the main variables and their interactions. This validation process enhances the implementation of ADoE and CCD by focusing on the most impactful factors affecting the injection moulding process.

The eight parameters were: mould temperature, melt temperature, holding time, holding pressure, shot size, switch-over position, injection speed, and cooling time. A two-level fractional factorial design was used for the initial screening analysis. The variables with their upper and lower limits are shown in Table 2.

The number of experiments for the fractional factorial method with a 1/16 fraction and eight two-level variables was 16. Each experiment design was repeated ten times (ten manufactured parts) to ensure stability, preserve data for part quality evaluation via metrology and to capture natural process variation during the process. To calculate the temperature differential ($\Delta T$ ) in each run, the average of each temperature sensor reading was calculated during the ten cycles and then the difference between the two averages was determined as per Eq. (9). The thermocouples used had a low error of $\pm 0.5^\circ C$, resulting in minimal variation in the measurement readings across the ten shots.

The variables having a significant effect on $\Delta T$ were selected from all eight variables to use in the ADoE and CCD experimental designs based on the screening results. A full central composite design was then applied using the process variables obtained in the screening section with the aim of having an accurate quadratic model and investigating the interaction between the variables. The design optimiser was used to find the optimum settings for the $\Delta T$ and cycle time by using the desirability function in Eq. (7).

## ADoE

The adaptive DoE required an initial sample data set, which is used to predict the next experimental designs. For the initial dataset, twelve experiments were randomly chosen from the CCD. The software package used in this paper for BO is "mlrMBO"[51]. It has the advantages of providing a good range of acquisition functions and it is suitable for multi-objective problems. Gaussian process model was chosen as the surrogate model, and the acquisition function was EI (expected improvement).

The data from the initial twelve experiments was used as prior data for the Bayesian optimisation algorithm to predict the design of the next two new experiments. Since the responses were collected inline during the process optimisation, the next two experiment settings were generated in the ADoE approach to save time in computing the next settings between every experiment. These two experimental designs were conducted for ten shots in the injection moulding machine to ensure the stability of the process and provide data for metrology.

The $\Delta T$ was measured using the sensor data Eq. (9), and the cycle time was also recorded. Then these two new designs were added to the initial dataset and fed into the ADoE to predict the next two experiment designs. This was repeated until it was observed that the recommended process settings had converged.

The criterion for the stopping point was stability or convergence of the input parameters for single objective optimisation. The stopping point for the multiobjective optimisation was defined by reaching the minimum thresholds set for the objectives. The thresholds were $7^\circ C$ on $\Delta T$ and 33s on cycle time. If any solution found during the BO meets or exceeds the threshold for the objectives, the BO can be considered successful. Figure 6 compares the optimisation process using the Bayesian adaptive design of experiment with the desirability function and NSGA-II.

## Optimisation with desirability function and NSGA-II

For modelling the relationship between the input parameters and each of the objectives ( $\Delta T$ and cycle time) different models were investigated using the CCD dataset; Artificial Neural Network (ANN) and Gaussian process (GP) modelling with fivefold cross-validation and the quadratic RSM model obtained from CCD. The ANN model utilised a feedforward architecture featuring ten neurons and Rectified Linear Unit (ReLU) activation function. The GP model employed a Gaussian kernel, and the hyperparameters, including the variance of the Gaussian process and the length scale parameter, were optimised through maximum likelihood estimation (ML). Table 3 summarises the RMSE of each method for each of the outputs. From the Table RSM model outperformed the other two methods since it has a lower RMSE for both cycle time and $\Delta T$. It is also worth noting that the ANN parameters and architecture were not optimised, so the ANN may achieve better performance compared to the other methods once optimisation is applied.

After finding a suitable model for each of the objectives, NSGA-II was applied to search for the optimal outputs based on the flowchart in Fig. 2. The algorithm was initiated with a population size and maximum iteration of 100 and a crossover probability of 0.8. The RSM model was also integrated with the desirability function to find the optimal process settings, first with $\Delta T$ as the response variable and then for two response variables: $\Delta T$ and cycle time.

In this paper, both single-objective and multi-objective optimisation were employed to address different goals:

| Level | Holding Time (s) | Mould temperature (°C) | Holding pressure (bar) | Shot size (mm) | Switch over position (mm) | Injection velocity (mm/s) | Barrel temperature (°C) | Cooling time (s) |
|---|---|---|---|---|---|---|---|---|
| Low | 3 | 40 | 200 | 6 | 5 | 20 | 210 | 10 |
| High | 6 | 90 | 400 | 12 | 10 | 40 | 230 | 30 |

**Table 2.** Selected process parameters and levels for screening.

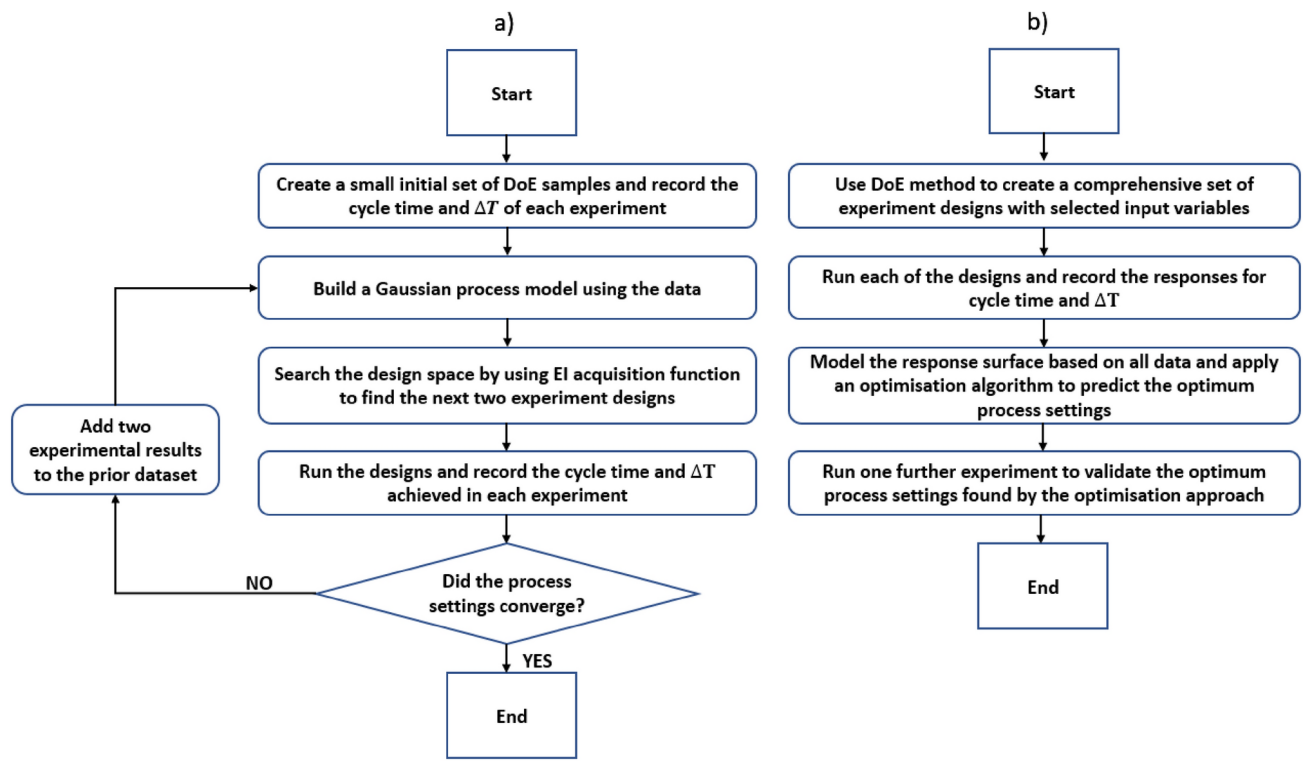

**Fig. 6.** (**a**) Bayesian ADoE and (b) desirability function & NSGA-II optimisation flowcharts.

| Model | Average RMSE $\Delta T$ (°C) | Average RMSE cycle time (s) |
|---|---|---|
| ANN | 0.68 | 2.36 |
| GP | 0.32 | 1.02 |
| RSM | 0.21 | 0.95 |

**Table 3.** Comparing ANN, GP, and RSM models.

1. Single-objective optimisation: The desirability function was used to optimise the temperature differential alone ($\Delta T$), which was a key factor correlated with quality issues like shrinkage. This step was essential for obtaining the best result for this specific parameter and enabled comparison with the ADoE results.

2. Multi-objective optimisation: Both the desirability function and NSGA-II were applied to minimise both cycle time and ($\Delta T$). NSGA-II allowed for finding Pareto-optimal solutions, offering a broader exploration of trade-offs between the objectives, and providing flexibility in optimising multiple competing factors. The results were compared to ADoE to assess and highlight ADoE's performance. The desirability function can still find an optimum solution for multi-objective optimisation by converting multiple objectives into a single composite score. However, this process inherently balances the objectives according to pre-defined importance weights or priorities, which might mask the true trade-offs between them. In contrast, NSGA-II used for comparison, preserves the trade-off information, offering a set of Pareto-optimal solutions where no objective can be improved without worsening another. Thus, the desirability function provides a solution but may not fully explore the balance between competing objectives.

## Metrology

After conducting the experiments, an OGP Smartscope CNC 500 dimensional multi-sensor coordinate measuring machine (CMM) with a resolution of $0.5\mu m$ was used for evaluating dimensional measurements. The physical measurements were taken at least two weeks after the production of the components to ensure no further variation in the dimensions on the relaxation of residual stresses. The CMM measurements were used to correlate the process settings with maximum, minimum, and average temperature differential ($\Delta T$) and verify the relationship between $\Delta T$ and shrinkage. Three main or critical dimensions shown in Fig. 4 (L, H, and W) were chosen for dimensional evaluation with a tolerance of $\pm 0.2$ mm, specified by the industrial partner. Figure 7 illustrates the CMM machine measuring ten clip components on a designed fixture. Based on this validation, the component with the maximum $\Delta T$ and the minimum $\Delta T$ should have the maximum and minimum shrinkage, respectively.

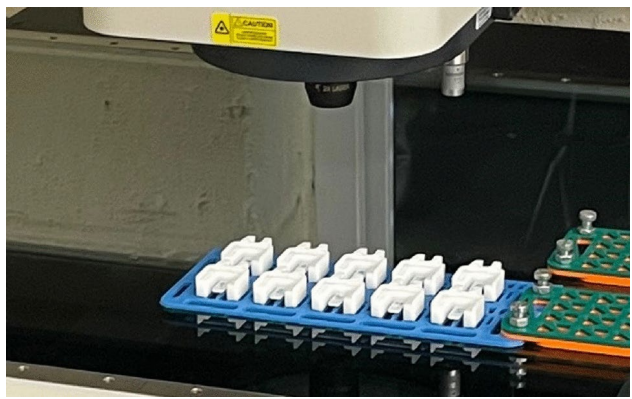

**Fig. 7.** OGP Smartscope CNC 500 measuring the main dimensions of Clip.

| No. | Run type | Run number | Actual $\Delta T$ (°C) | $\Delta T$ - type | Cycle time (s) | Tolerance deviation (standard deviation) | | |
|---|---|---|---|---|---|---|---|---|
| | | | | | | L (%) | H (%) | w (%) |
| 1 | ADoE-multiobjective | 4 | 10.19 | Maximum | 29.8 | 150.4 (22.8) | 50.75 (20.12) | 152.25 (7.8) |
| 2 | ADoE-multiobjective | 9 | 7.77 | Average | 28.8 | 137.8 (4.4) | 47.75 (21.47) | 142.75 (4.6) |
| 3 | CCD | 23 | 7.63 | Average | 33.3 | 77 (5.04) | 35.3 (25) | 122.75 (9.4) |
| 4 | Multiobjective desirability function optimum setting | – | 7.42 | Average | 28.8 | 57.4 (6.8) | 48.8 (36.5) | 90.8 (3.6) |
| 5 | ADoE-multiobjective optimum setting | 10 | 6.51 | Minimum | 32.9 | 52 (3.8) | 42 (16.15) | 85.6 (5.44) |
| 6 | ADoE & desirability function- single-optimum setting | 4 | 5.54 | Minimum | 44.8 | 46.6 (1.03) | 24.75 (15.19) | 78.5 (4.24) |

**Table 4.** Metrology results, presented in descending order of $\Delta T$ recorded in-process.

## Results and discussion

### Metrology

Parts produced under various process settings were selected for metrology investigation using both CCD and optimisation approaches. This was done to evaluate the correlation between $\Delta T$ and post-production shrinkage. The experiments were designed to encompass a range of $\Delta T$ values, from 5.5 to 10.19 °C, to ensure a comprehensive analysis. The selected experiments are listed in Table 4, and $\Delta T$ is presented in descending order for clarity. In Table 4, the tolerances are defined as percentages by considering 100% equal to the specified product tolerance of ±0.2 mm. Each main dimension with a value less than 100% is in an acceptable range, meaning post-production shrinkage has not significantly occurred in the component. For each experiment, the tolerances of ten components were measured and the average value was calculated and noted in Table 4 with their corresponding standard deviations.

For better visualisation, Fig. 8 summarises Table 4 into a bar graph and illustrates the tolerance deviation from the baseline of 100% (±0.2 mm) for each of the dimensions in each experiment. Based on Table 4 and Fig. 8, the reduction in $\Delta T$ led to a decrease in tolerance variation of each critical dimension, and for the optimum process settings found by ADoE, no shrinkage was observed in any dimension. Hence, this correlation validates that $\Delta T$ can be considered as the shrinkage index to be used in real-time in the ADoE approach. Incorporating shrinkage as the objective necessitates waiting for post-production shrinkage, a process that takes up to several weeks as the residual stresses gradually relax over time. In contrast, optimising based on $\Delta T$ enables inline and rapid adjustments.

### Screening and CCD step

The effect of eight input parameters on the $\Delta T$ was investigated using fractional factorial design in Minitab 20. A 95% confidence bound was selected, such that any variables with a p-value lower than 0.05 were deemed significant.

Figure 9a, b show the normal plots and Pareto chart of the effects from analysis of variance (ANOVA). Based on Fig. 9, the mould temperature, cooling time, holding time, barrel (melt) temperature, the interaction between cooling time and holding time, and the switch-over position affected the $\Delta T$.

The parameters identified in this study align with the findings of previous research on effective process parameters on warpage in the IM process. For example, previous studies on plastic product warpage and shrinkage[7,52–54] identified melt temperature, mould temperature, packing time, cooling time, and switch-over position as the most influential process parameters on shrinkage and warpage. It is essential to emphasise that process parameter screening for cycle time was not conducted since the only influential process parameters on cycle time are packing time and cooling time, both were considered as process inputs in the optimisation problem.

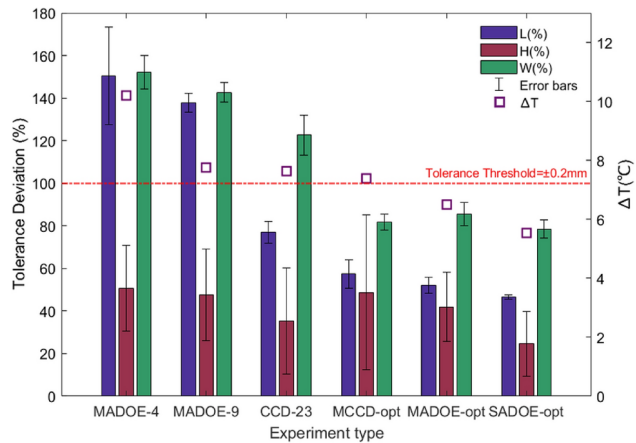

**Fig. 8.** Bar graphs of the tolerance deviation of the main dimensions in Clip. MADOE refers to multiobjective ADOE and for the examination of metrology in MADOE experiment numbers 4, 9 and the optimum setting found by ADOE were selected. MCCD-opt refers to multiobjective optimisation using CCD and the desirability function. SADOE means the single objective optimisation of $\Delta T$ using ADOE.

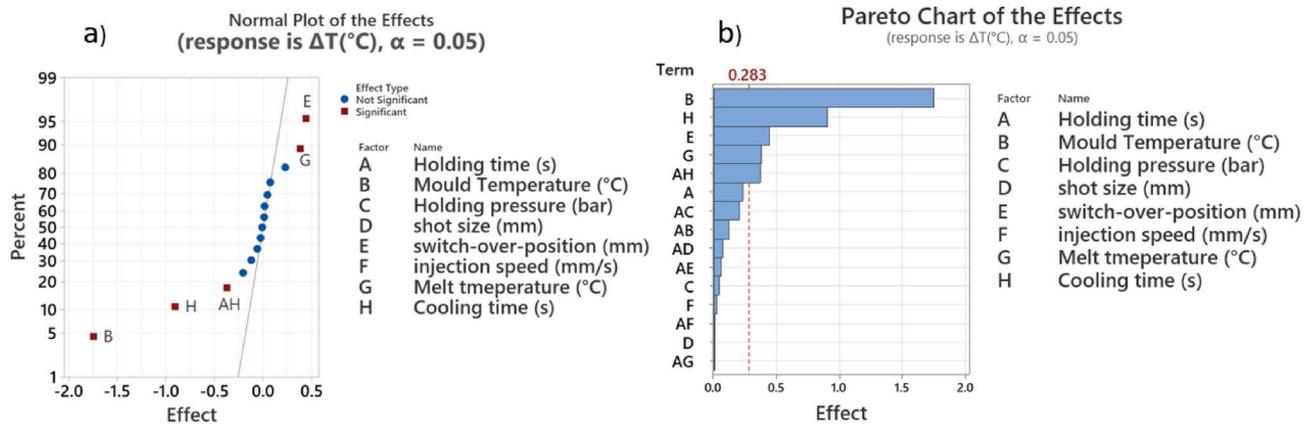

**Fig. 9.** (**a**) Normal plots and (**b**) Pareto chart of ANOVA for screening.

| Parameters | Level | | | | |
|---|---|---|---|---|---|
| | $-\alpha$ | $-1$ | $0$ | $1$ | $+\alpha$ |
| Mould temperature ($^\circ$C) | 55 | 65 | 75 | 85 | 95 |
| Cooling time (s) | 15 | 20 | 25 | 30 | 35 |
| Holding time (s) | 1.5 | 3 | 4.5 | 6 | 7.5 |
| Barrel temperature ($^\circ$C) | 195 | 205 | 215 | 225 | 235 |

**Table 5.** Selected process parameters and levels for CCD.

Based on screening results, four variables were selected for the optimisation to use in CCD and ADoE. These were: mould temperature, cooling time, holding time, and barrel (melt) temperature, as detailed with their respective levels in Table 5. The switch-over position depends on the shot size and leftover melt cushion; this is challenging to modify during the injection moulding machine cycles. Although it was identified as an influential design variable in ANOVA, it was removed from the main process parameters for this reason. The selected CCD method in this study is circumscribed (CCC), and the $\alpha$ value is equal to 2. The number of experiments for a full CCD design, using four input variables ($f = 4$) and six replications at the centre point ($r = 6$), is 31. The experiments were designed using Minitab 20 software. Table 6 shows the experimental designs for the CCD and the assigned values for the cycle time and average $\Delta T$ for each experiment.

The normal plots and Pareto charts for the CCD design, presented in Fig. 10, validate the selection of the variables from the fractional factorial design since all four selected variables have a significant effect on the average $\Delta T$. ANOVA was used for average $\Delta T$ to evaluate the predicted model by CCD. The results are summarised in Table 7, showing the p-values of the parameters and the coefficient of prediction (R-squared predicted), which is higher than 90%.

| Experiment No | Mould Temperature ($^\circ$C) | Cooling Time (s) | Holding time (s) | Barrel temperature ($^\circ$C) | Avg $\Delta T$ ($^\circ$C) | Cycle time (s) |
|---|---|---|---|---|---|---|
| 1 | 65 | 30 | 3 | 205 | 7.6 | 40.3 |
| 2 | 65 | 20 | 6 | 205 | 8.84 | 33.3 |
| 3 | 75 | 25 | 4.5 | 215 | 8.12 | 36.8 |
| 4 | 75 | 25 | 4.5 | 215 | 8.22 | 36.8 |
| 5 | 65 | 20 | 3 | 205 | 9.8 | 30.3 |
| 6 | 75 | 25 | 4.5 | 215 | 8.1 | 36.8 |
| 7 | 65 | 20 | 6 | 225 | 9.46 | 33.3 |
| 8 | 85 | 30 | 6 | 225 | 6.82 | 43.3 |
| 9 | 75 | 25 | 4.5 | 215 | 8.17 | 36.8 |
| 10 | 75 | 25 | 4.5 | 215 | 8.17 | 36.8 |
| 11 | 65 | 30 | 3 | 225 | 8.1 | 40.3 |
| 12 | 75 | 35 | 4.5 | 215 | 6.78 | 46.8 |
| 13 | 95 | 25 | 4.5 | 215 | 7.1 | 36.8 |
| 14 | 85 | 30 | 6 | 205 | 6.5 | 43.3 |
| 15 | 75 | 25 | 1.5 | 215 | 8.23 | 33.8 |
| 16 | 55 | 25 | 4.5 | 215 | 8.74 | 36.8 |
| 17 | 75 | 15 | 4.5 | 215 | 9.62 | 26.8 |
| 18 | 75 | 25 | 4.5 | 235 | 8.4 | 36.8 |
| 19 | 75 | 25 | 7.5 | 215 | 7.63 | 39.8 |
| 20 | 85 | 30 | 3 | 205 | 6.66 | 40.3 |
| 21 | 75 | 25 | 4.5 | 215 | 8.1 | 36.8 |
| 22 | 85 | 20 | 6 | 225 | 8.4 | 33.3 |
| 23 | 85 | 20 | 6 | 205 | 7.63 | 33.3 |
| 24 | 65 | 30 | 6 | 205 | 6.97 | 43.3 |
| 25 | 85 | 20 | 3 | 225 | 8.37 | 30.3 |
| 26 | 65 | 20 | 3 | 225 | 9.34 | 30.3 |
| 27 | 65 | 30 | 6 | 225 | 8.27 | 43.3 |
| 28 | 85 | 30 | 3 | 225 | 6.94 | 40.3 |
| 29 | 85 | 20 | 3 | 205 | 8.03 | 30.3 |
| 30 | 75 | 25 | 4.5 | 195 | 7.7 | 36.8 |
| 31 | 75 | 25 | 4.5 | 215 | 7.94 | 36.8 |

**Table 6.** CCD design and results for the average $\Delta T$ and cycle time.

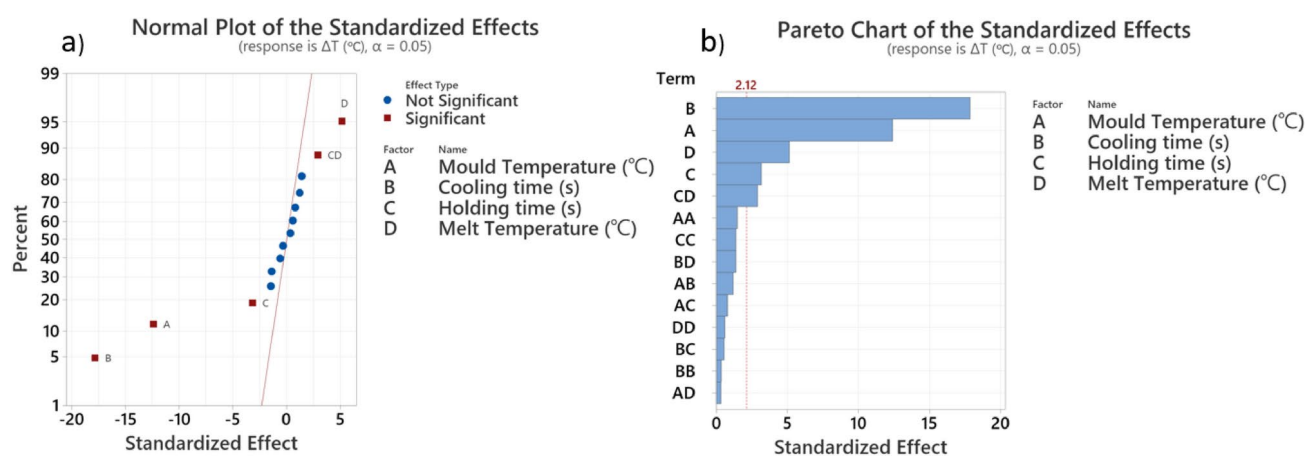

**Fig. 10.** (**a**) Normal plots and (**b**) Pareto chart of ANOVA for CCD.

| P-value | | | | | | Model | | |
|---|---|---|---|---|---|---|---|---|
| Model | Mould temperature | Cooling time | Holding time | Barrel temperature | Holding time * barrel temperature | R-sq% | R- sq(adj)% | R-sq(pred)% |
| < 0.005 | < 0.005 | < 0.005 | 0.004 | < 0.005 | 0.007 | 95.43 | 94.51 | 92.24 |

**Table 7.** CCD model ANOVA results for $\Delta T$.

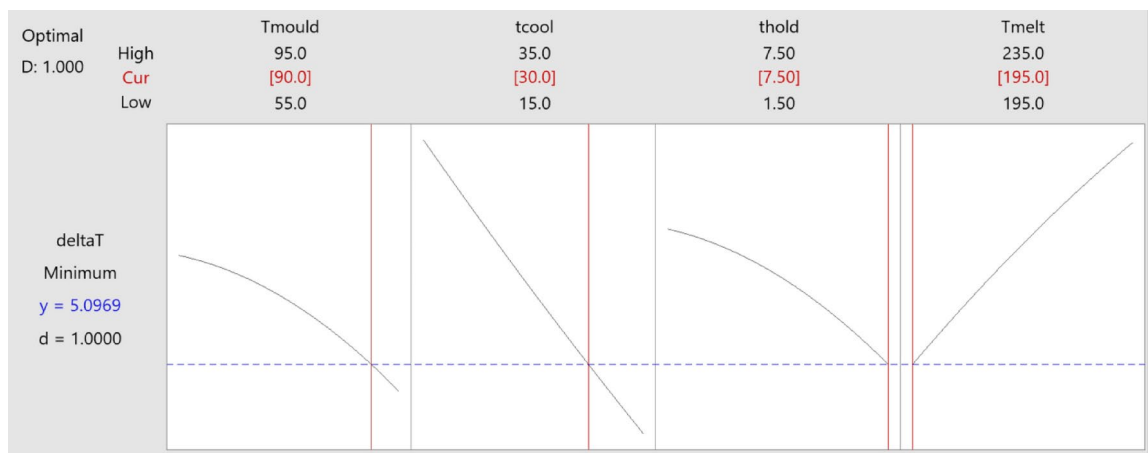

**Fig. 11.** Optimisation by desirability function for single response $\Delta T$.

| Mould temperature (°C) | Cooling time (s) | Holding time (s) | Barrel temperature (°C) | Avg $\Delta T$ (°C) |
|---|---|---|---|---|
| 90 | 30 | 7.5 | 195 | 5.1 |

**Table 8.** Optimum process settings found by desirability function for $\Delta T$.

| Mould temperature (°C) | Cooling time (s) | Holding time (s) | Barrel temperature (°C) | Avg T (°C) | Cycle time (s) |
|---|---|---|---|---|---|
| 90 | 30 | 7.5 | 195 | 5.1 | 30.28 |

**Table 9.** Optimum process settings found by desirability function for multi-response optimisation.

Based on the significant process parameters and their levels listed in Table 5, the mathematical multi-objective optimisation problem can be formulated as follows:

$$\min F(x) = [\Delta T(x), \text{cycle time}(x)] \tag{10}$$

where $x = $ [mould temperature, cooling time, holding time, barrel temperature]. The objective is to minimise the temperature differential ($\Delta T$) between a part's hotspot and average temperature, as well as the cycle time, by optimising the design variables represented by $x$.

### Optimisation by using the desirability function and RSM model

The design optimiser in Minitab 20 was applied to use the desirability function, and the result for $\Delta T$ is illustrated in Fig. 11. Based on this result, the optimum process setting to minimise the $\Delta T$, at a value of just under 5.1°C, is summarised in Table 8. The desirability value for this single response is equal to 1 based on Eqs. (7) and (8).

After single-response optimisation, a multi-objective optimisation was performed using the desirability function in Minitab 20. The result for the optimum process settings to provide a process with the minimum cycle time and $\Delta T$ is summarised in Table 9 and Fig. 12. In Fig. 12, the vertical red lines indicate the optimal value for each of the process settings, while the two horizontal blue lines represent the optimal cycle time and $\Delta T$ that can be achieved based on these settings. Based on this result, using the optimum settings in Table 9 and Fig. 12, $\Delta T$ will be 7.42°C and the cycle time 30.28 s. The individual desirability value for $\Delta T$ $d_1(y_1)$ is equal to 0.72102 and for cycle time $d_2(y_2)$ is 0.82617. So, the composite desirability based on Eq. (7) with n = 2 will be 0.7718. The weight (*w*) of each response was considered as 1 in Eq. (7).

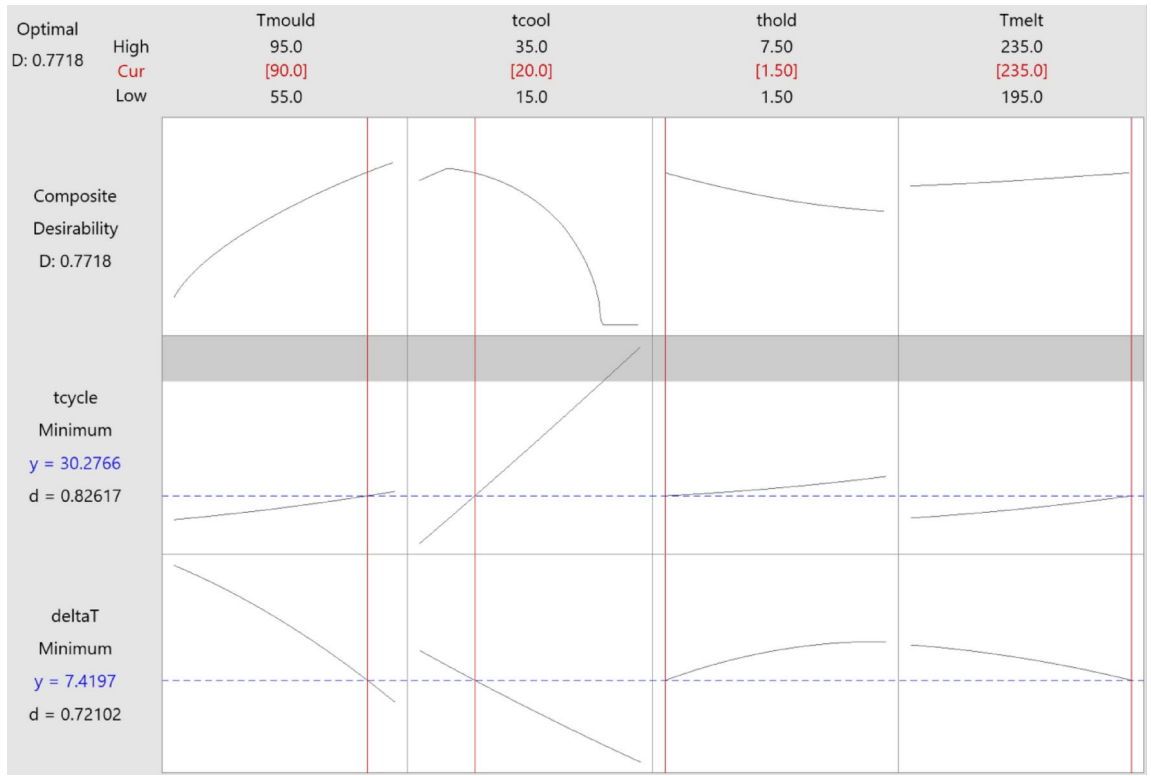

**Fig. 12**. Optimisation by desirability function for multi-response $\Delta T$ and cycle time.

| No | Mould temperature (°C) | Cooling time (s) | Holding time (s) | Barrel temperature (°C) | $\Delta T$ (°C) |
|----|------------------------|------------------|------------------|-------------------------|-----------------|
| 13 | 90.5                   | 30               | 7.2              | 195                     | 5.85            |
| 14 | 93.95                  | 30               | 6.8              | 195                     | 5.84            |
| 15 | 90                     | 30               | 7.5              | 195.5                   | 5.54            |
| 16 | 90                     | 30               | 7.5              | 195.5                   | 5.54            |

**Table 10**. Optimum process settings found by ADoE for average $\Delta T$ after four experiments.

## ADoE

For the comparison of the adaptive DoE in finding the optimum settings with desirability function, the same testing strategy was applied. The ADoE was first carried out for the $\Delta T$ and then for the $\Delta T$ and cycle time as the multi-objective optimisation approach.

The twelve experiments were randomly selected from the CCD as prior data for the Bayesian optimisation, and the optimisation process was followed based on the flowchart in Fig. 6. The optimum setting was found for the $\Delta T$ after just four more experiments (or two loops in the flowchart), and the experimental settings converged based on the stopping criterion outlined in the methodology section for single response optimisation. The extra four experiments are listed in Table 10. A total of 16 experiments were required to find the optimum setting for the single optimisation of $\Delta T$.

For the multi-objective ADoE, a similar approach was followed based on Fig. 6. After conducting ten more experiments using the Bayesian optimisation approach (5 loops of the flowchart) in addition to the initial 12 experiments, a desired value for the $\Delta T$ and cycle time were found, 6.51°C and 32.9s respectively (Table 11). This was based on the stopping criterion mentioned in section 3.3 for multi-response optimisation. For the multi-objective design, a total of 22 experiments were conducted, which is more efficient than the initial CCD experiments with 31 designs.

## NSGA-II

The RSM models for $\Delta T$ and cycle time were derived utilising CCD data. These models were then employed as objective functions in the NSGA-II approach to determine the optimal process settings aimed at minimising both $\Delta T$ and cycle time. The lower and upper bounds of input parameters were defined in Table 5.

| No. | Mould temperature (°C) | Cooling time (s) | Holding time (s) | Barrel temperature (°C) | $\Delta T$ (°C) | Cycle time (s) |
|---|---|---|---|---|---|---|
| 1 | 90 | 30 | 2.3 | 195 | 6.3 | 39.6 |
| 2 | 88.6 | 24.8 | 4.5 | 200.7 | 6.96 | 36.6 |
| 3 | 90 | 30 | 7.5 | 195.5 | 5.54 | 44.8 |
| 4 | 70 | 20.2 | 2.35 | 231.5 | 10.19 | 29.8 |
| 5 | 87.1 | 24.8 | 1.92 | 199.5 | 7.21 | 34.02 |
| 6 | 90 | 28.3 | 1.75 | 195.2 | 6.75 | 37.35 |
| 7 | 90 | 30 | 4.5 | 195.3 | 6.16 | 41.1 |
| 8 | 90 | 25 | 1.5 | 195 | 6.8 | 33.8 |
| 9 | 90 | 20 | 1.5 | 195 | 7.77 | 28.8 |
| 10 | 90 | 24.1 | 1.5 | 195.1 | 6.51 | 32.9 |

**Table 11.** Optimum process settings found by ADoE for average $\Delta T$ and cycle time after ten experiments.

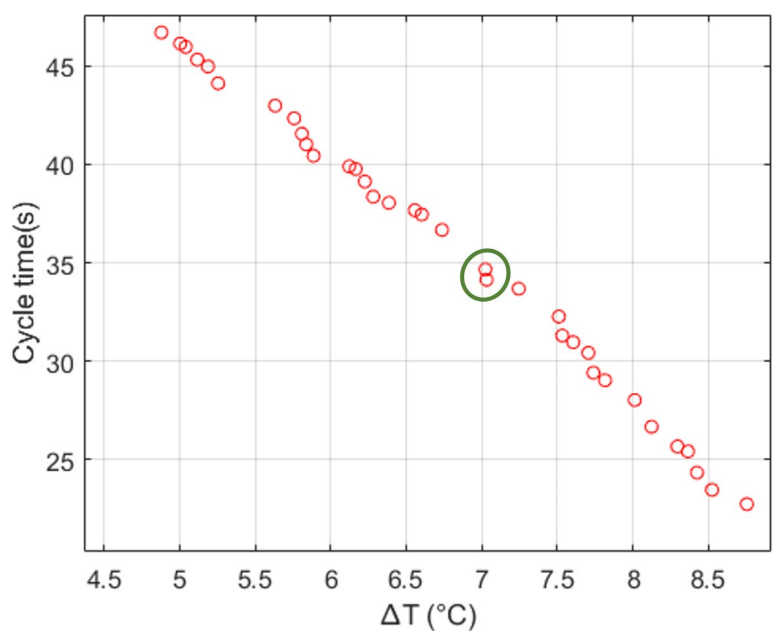

**Fig. 13.** Pareto fronts of NSGA-II optimisation using RSM model to minimise cycle time and $\Delta T$.

| Mould temperature (°C) | Cooling time (s) | Holding time (s) | Barrel temperature (°C) | Predicted $\Delta T$ (°C) | Predicted cycle time (s) |
|---|---|---|---|---|---|
| 88.3 | 1.63 | 26.24 | 198.48 | 7.03 | 34.14 |

**Table 12.** NSGA-II optimal Pareto front.

Following Algorithm 2, thirty-five optimal Pareto fronts were identified using NSGA-II, as shown in Fig. 13. This reflects that, a longer cycle time will result in a smaller $\Delta T$, as expected. Based on the desired threshold defined for the responses, $\Delta T \leq 7$°C and cycle time $\leq 33$s, there are only two appropriate solutions as shown with a green circle in the figure. Between these two solutions, the one with a lower cycle time is more desirable. Table 12 summarises the input parameters and responses for this selected Pareto optimal solution. This solution resulted in a cycle time of 34.14 s and $\Delta T$ of 7.03°C.

## Comparison of optimisation with ADoE approach, desirability function and NSGA-II

The result of each method is summarised in Table 13. Optimisation with desirability function and NSGA-II requires a further additional experiment to test the predicted optimum values, while ADoE is based on the experimental results when convergence to the desired performance is known to be achieved. Both the predicted and actual responses for the optimisation methods are outlined in the table. The first advantage of ADoE over the other optimisation approaches, presented in Table 13, is that it uses significantly fewer experiments for optimisation. The number of experiments for the single optimisation of $\Delta T$ was almost halved, from 31 experiments for other methods to 16 for ADoE. For the multi-response optimisation, the number of experiments for ADoE was reduced to 22 compared to 31 for the desirability function and NSGA-II. The predicted optimum process settings for the single optimisation of $\Delta T$ applying desirability function and ADoE are similar, as indicated in Table 13. Since ADoE results are based on the experimental data, a good agreement between the experimental results ($\Delta T = 5.54$°C) and optimisation using the desirability function ($\Delta T = 5.10$°C) can be observed. The predicted optimum process settings for multi-response optimisation using the desirability function were also conducted in experiments for verification. The experimental values for $\Delta T$ and cycle time were 8.2°C, and 28.8s, and the estimated values by desirability function based on Table 13 were 7.42°C and 30.28 s. The errors in cycle time and $\Delta T$ prediction for the desirability function were almost 5% and 10%, respectively.

| Optimisation method | No of experiments | Mould temperature (°C) | Cooling time (s) | Holding time (s) | Barrel temperature (°C) | $\Delta T$ (°C) | | Cycle time (s) | |
|---|---|---|---|---|---|---|---|---|---|
| | | | | | | Predicted | Actual | Predicted | Actual |
| Single desirability function | 31 | 90 | 30 | 7.5 | 195 | 5.1 | 5.54 | - | - |
| Single ADoE | 16 | 90 | 30 | 7.5 | 195.5 | - | 5.54 | - | - |
| multi- desirability function | 31 | 90 | 20 | 1.5 | 235 | 7.42 | 8.2 | 30.28 | 28.8 |
| multi ADoE | 22 | 90 | 24.1 | 1.5 | 195.1 | - | 6.51 | - | 32.9 |
| NSGA-II | 31 | 89 | 26 | 1.63 | 198.5 | 7.04 | 6.83 | 34.14 | 34.93 |

**Table 13.** Optimum process parameters found by different optimisation methods for $\Delta T$ and cycle time.

From Table 13, it is also evident that for the optimisation of the $\Delta T$, both methods found similar process settings. However, for the multi-optimisation of the cycle time and $\Delta T$, the barrel temperature is 235°C for the desirability function and 195.1°C for ADoE. The temperature of 235°C is relatively high for the POM material and causes other defects in the part, such as burn marks. The ADoE framework with a Gaussian process trained on 12 initial experiments was successful after four experiments (repeating the flowchart of Fig. 6 twice) and determined that for optimising the $\Delta T$ and cycle time together, the barrel temperature should stay around 195°C.

The NSGA-II optimal process settings closely mirror those for ADoE optimal settings in multiobjective optimisation, differing mainly in a longer cooling time of approximately 2 s. The average RMSE values, approximately 1 s and 0.21°C for cycle time and $\Delta T$, suggest that this predicted optimal setting aligns closely with the experimental one obtained through ADoE. However, it took a total number of 31 experiments to build the RSM models and apply the optimisation algorithm for NSGA-II, along with one additional experiment for validation. In contrast, ADOE required only 22 experiments to identify the optimal responses. Furthermore, the ADoE-based yielded a smaller $\Delta T$ and a shorter cycle time, both of which are more desirable outcomes.

The correlation between dimensional stability and $\Delta T$ is clear in Table 4 and Fig. 8. The height (H) is always within the defined tolerance boundary for any $\Delta T$ value, but the tolerances for length (L) and width (W) decrease considerably as $\Delta T$ decreases. For example, a reduction in $\Delta T$ from 10.19°C to 5.54°C corresponds to a significant decrease in tolerance error for the length of the component (L), from 150.4% (−0.3 mm) to 46.6% (−0.093 mm). The optimum process settings found by ADoE resulted in parts meeting the dimensional tolerances. $\Delta T$ values below 7.5°C resulted in components satisfying the dimensional tolerance requirement. However small increases in $\Delta T$ above this threshold corresponded in one or more dimensions being out of tolerance and therefore a lower target range of $\Delta T$ from 6.5°C to 7°C is recommended to ensure the resulting components will meet specification. In the future, more data on the relationship between $\Delta T$ and the dimensional stability of the parts could help to tune the optimum threshold more exactly, to ensure the required part quality without extending the cooling time any longer than necessary.

One further point to mention is that the initial dataset for the ADoE method included data from twelve randomly selected CCD runs. Using a smaller or different baseline dataset can affect the ADoE loop in Fig. 6, making it longer or shorter depending on the selected data. However, irrespective of the starting dataset, the ADoE approach can be expected to converge on the same optimum process settings for minimising the cycle time and temperature differential. In this research, the optimisation process for CCD and ADoE was done in less than two days by conducting more than sixty experiments. In contrast, even with a high-spec system, running the sixty simulations is significantly time-consuming and typically requires several days as well as expertise with expensive simulation software.

## Conclusion and future work

In this research, a Bayesian adaptive design of experiment method (ADoE) was used in an experimental IM process to optimise product quality and cycle time. In ADoE , it is crucial to quickly assess the real-time response of the component. Therefore, the desired outputs included the cycle time, recorded directly from the machine, and the in-line temperature differential ($\Delta T$), which was manually calculated using data from two thermocouples. In this proposed approach for prediction of the product quality by ADoE and using $\Delta T$ , only one simulation was conducted in Moldflow to identify the hotspot locations and place the sensors. The inline $\Delta T$ was used as a predictor of dimensional stability and shrinkage verified by metrology. A classic DOE approach, full central composite design (CCD), combined with desirability function and NSGA-II algorithm was also implemented for comparison to ADoE. Injection moulding experiments were conducted on an industrial-standard machine to verify and compare the methods using the same moulded component. The ADoE approach improves significantly on the standard other approaches in both single and multi-objective optimisation, in terms of efficiency (no of experiments) and the performance of the resulting recommended settings.

The Bayesian adaptive design applies prior knowledge to iterate toward the optimum process settings. Also, since the Bayesian adaptive DoE implements both exploitation (where the objective mean is high) and exploration (where the uncertainty is high) for searching, it requires a smaller dataset, offering better performance than CCD. This study implemented ADoE in an actual injection moulding machine and validated the efficiency of ADOE in an actual manufacturing process using experimental and inline data rather than simulation data[36,39]. Not only is the experimental approach more reliable, it is also faster than running an equivalent number of simulations. It has the further advantage that it can be quickly adapted to changing raw material batch or other changes to machine set-up.

ADoE requires a smaller dataset compared to the classic DoE methods that are used for optimisation of injection moulding processes. Since running experiments can be expensive, resource-intensive, and time-consuming, ADoE can reduce the optimisation cost, scrap rates, and enhance the overall component quality by decreasing defects associated with warpage and shrinkage. The proposed approach was applied for two responses; however, the approach could be extended to consider other responses such as energy or material consumption. Future studies could compare this type of adaptive Design of Experiments (ADoE) approach to other adaptive DoE methods, utilising different initial experimental sets, surrogate models, and acquisition functions, while incorporating a broader range of response variables. Additionally, future work could benchmark results against other advanced multi-objective optimisation methods, such as improved dragonfly optimisation[55] and other cutting-edge techniques, to assess comparative performance and robustness.

## Data availability
All data used in this study are included in this published article.

## Acknowledgements

This publication has emanated from research supported by an ATU Sligo Bursary and a grant from Science Foundation Ireland under Grant number 16/RC/3872. For the purpose of Open Access, the author has applied a CC BY public copyright licence to any Author Accepted Manuscript version arising from this submission.

## Author contributions

M.K.: Conceptualisation, Methodology, Software, Validation, Formal Analysis, Investigation, Data Curation, Visualisation, Writing - original draft, Writing - review & editing; D.T.: Funding acquisition, Project administration, Conceptualisation, Methodology, Investigation, Supervision, Resources, Writing- review & editing; C.R.: Investigation, Writing - review & editing; C.O.: Methodology, Visualisation, Writing - review & editing; A.W.: Methodology, Visualisation, Writing - review& editing; M.M.: Funding acquisition, Project administration, Conceptualisation, Methodology, Validation, Formal Analysis, Supervision, Writing - review & editing.

## Competing interests

The authors declare no competing interests.

## Additional information

**Correspondence** and requests for materials should be addressed to M.M.

**Reprints and permissions information** is available at www.nature.com/reprints.

**Publisher's note**  Springer Nature remains neutral with regard to jurisdictional claims in published maps and institutional affiliations.

**Open Access**   This article is licensed under a Creative Commons Attribution 4.0 International License, which permits use, sharing, adaptation, distribution and reproduction in any medium or format, as long as you give appropriate credit to the original author(s) and the source, provide a link to the Creative Commons licence, and indicate if changes were made. The images or other third party material in this article are included in the article's Creative Commons licence, unless indicated otherwise in a credit line to the material. If material is not included in the article's Creative Commons licence and your intended use is not permitted by statutory regulation or exceeds the permitted use, you will need to obtain permission directly from the copyright holder. To view a copy of this licence, visit http://creativecommons.org/licenses/by/4.0/.

© The Author(s) 2024